\documentclass[prl,twocolumn]{revtex4}
\usepackage{graphicx}

\newcommand{\be}{\begin{equation}}

\newcommand{\ee}{\end{equation}}
\newcommand{\bea}{\begin{eqnarray}}
\newcommand{\eea}{\end{eqnarray}}
\newcommand{\bse}{\begin{subequations}}
\newcommand{\ese}{\end{subequations}}

\begin{document}

\def\K{{{K}}}
\def\Q{{{Q}}}
\def\Gbar{\bar{G}}
\def\tk{\tilde{\bf{k}}}
\def\k{{{k}}}
\def\q{{\bf{q}}}

\title{Proximity of the Superconducting Dome and the Quantum Critical Point in the Two-Dimensional Hubbard Model}

\author{S.-X.\ Yang$^{1}$, H.\ Fotso$^{1}$, S.-Q.\ Su$^{1,2,*}$, D.\ Galanakis$^{1}$, E.\ Khatami$^{3}$, J.-H. She$^{4}$, J. Moreno$^{1}$, J. Zaanen$^{4}$, and M.\ Jarrell$^{1}$}
\affiliation{$^{1}$Department of Physics and Astronomy, Louisiana State University, Baton Rouge, Louisiana 70803, USA\\
$^{2}$Computer Science and Mathematics Division,Center for Nanophase Materials Sciences, Oak Ridge National Laboratory, Oak Ridge, Tennessee 37831-6164, USA \\
$^{3}$Department of Physics, Georgetown University, Washington, District of Columbia, 20057, USA\\
$^{4}$ Instituut-Lorentz for Theoretical Physics, Universiteit Leiden, P.O. Box 9506, 2300 RA Leiden, The Netherlands \\
$^{*}$ shiquansu@hotmail.com}

\begin{abstract}

We use the dynamical cluster approximation to understand the proximity of the superconducting 
dome to the quantum critical point in the two-dimensional Hubbard model.  In a BCS formalism, 
$T_c$ may be enhanced through an increase in the $d$-wave pairing interaction ($V_d$) or the bare 
pairing susceptibility ($\chi_{0d}$).   At optimal doping, where $V_d$ is revealed to be featureless, 
we find a power-law behavior of $\chi_{0d}(\omega=0)$, replacing the BCS log, and strongly
enhanced  $T_c$.  We suggest experiments to verify our predictions.

\end{abstract}

%74.20.-z 	Theories and models of superconducting state
%74.20.Fg 	BCS theory and its development 
%74.25.Dw 	Superconductivity phase diagrams 
%71.10.-w 	Theories and models of many-electron systems

\maketitle

%==========BODY OF PAPER =========================================

%
\paragraph*{Introduction-}

The unusually high superconducting transition temperature of the cuprates
remains an unsolved puzzle, despite more than two decades of intense theoretical
and experimental research. Central to the efforts to unravel this mystery
is the idea that the high critical temperature is due to the presence of a
quantum critical point (QCP) which is hidden under the superconducting dome~\cite{QCPreview}.
Numerical calculations in the Hubbard model, which is accepted as the de-facto
model for the cuprates, strongly support the case of a finite-doping QCP
separating the low-doping region, found to be a non-Fermi liquid (NFL),
from a higher doping Fermi-liquid (FL) region~\cite{Vidhyadhiraja09,jarrell01}.
Calculations also show that in the vicinity of the QCP,
and for a wide range of temperatures, the doping and temperature  dependence of the
single-particle properties, such as the quasi-particle weight~\cite{Vidhyadhiraja09},
as well as thermodynamic properties such as the chemical potential and the entropy,
are consistent with marginal Fermi liquid (MFL) behavior~\cite{karlis}.  This QCP emerges by tuning 
the temperature of a second-order critical point of charge separation transitions to zero
and is therefore intimately connected to $ q=0$ charge fluctuations~\cite{ehsan}. Finally,
the critical doping seems to be in close proximity to the optimal doping for
superconductivity as found both in the context of the Hubbard~\cite{ehsan} and the
t-J model~\cite{Kotliar}. Even though this proximity may serve as an indication
that the QCP enhances pairing, the detailed mechanism is largely unknown.

In this Letter, we attempt to differentiate between two incompatible scenarios
for the role of the QCP in superconductivity. The {\em{first}} scenario is the
quantum critical BCS (QCBCS) formalism introduced by She and Zaanen (She-Zaanen)~\cite{Jan09}.
According to this, the presence of the QCP results in replacing the
logarithmic divergence of the BCS pairing bubble by an algebraic divergence.
This leads to a stronger pairing instability and higher critical temperature
compared to the BCS for the same pairing interactions. The {\em{second}} scenario
suggests that remnant fluctuations around the QCP mediate the pairing
interaction~\cite{rome,Moon}. In this case the strength of the pairing interaction would
be strongly enhanced in the vicinity of the QCP, leading to the superconducting
instability.
Here, we find that near the QCP, the pairing interaction depends monotonically on the 
doping, but the bare pairing susceptibility acquires an algebraic dependence on the temperature, 
consistent with the first scenario.
\paragraph*{Formalism-}
\label{sec:formalism}
In a conventional BCS superconductor, the superconducting transition temperature, $T_c$, is 
determined by the condition $V \chi_0^\prime (\omega=0)=1$, where $\chi_0^\prime$ is the real 
part of the $ q=0$ bare pairing susceptibility, and $V$ is the strength of the pairing 
interaction.  The transition is driven by the divergence of $\chi_0^\prime (\omega=0)$
which may be related to the imaginary part of the susceptibility via
$\chi_0'(\omega=0)=\frac{1}{\pi} \int d\omega \chi_0''(\omega)/\omega$. 
And $\chi_0''(\omega)$ itself can be related to the spectral function,
$A_{\k}(\omega)$, through 
\begin{equation}
\chi''_0(x) = \frac{\pi}{N} \sum_{\zeta,\k} \int d\omega A_{\k}(\omega) A_{\k}(\zeta x-\omega)\left( f(\omega-\zeta x)-f(\omega) \right)
\label{eq:chipp}
\end{equation}
where the summation of $\zeta\in\{-1,+1\}$ is used to anti-symmetrize $\chi_0''(\omega)$.
In a FL, $\chi_0''(\omega) \propto N(\omega/2)\tanh\left(\omega/4T\right)$, and 
$\chi_0^\prime (T) \propto N(0) \ln(\omega_D/T)$ with $N(0)$  the single-particle density of 
states at the Fermi surface and $\omega_D$ the phonon Debye cutoff frequency.  This yields the 
well known BCS equation $T_c=\omega_D \exp[- 1/(N(0) V)]$.  In the QCBCS formulation, 
the BCS equation is $V\chi'(\omega=0)=1$, where $\chi'$ is fully dressed by both the self energy 
and vertices associated with the interaction responsible for the QCP, but not by the 
pairing interaction $V$.  In the Hubbard model % as in the cuprates,
the Coulomb interaction is 
responsible for both the QCP and the pairing, so this deconstruction is not possible.  Thus, we 
will use the more common BCS $T_c$ condition to analyze our results with $V\chi_0'(\omega=0)=1$ 
where $\chi_0'$ is dressed by the self energy but without vertex corrections.  Since the QCP 
is associated with MFL behavior, we do not expect the bare bubble to display a FL
logarithm divergence.  Here, we explore the possibility that 
$\chi_0'(\omega=0) \sim 1/T^{\alpha}$.

The two-dimensional Hubbard model is expressed as:  
\begin{equation}
H=H_k+H_p=\sum_{\k\sigma}\epsilon_{\k}^{0}c_{{\k}\sigma}^{\dagger}c_{{\k}\sigma}^{\phantom{\dagger}}+U\sum_{i}n_{{i}\uparrow}n_{{i}\downarrow} \,,
\label{eq:hubbard}\end{equation}
 where $c_{{\k}\sigma}^{\dagger}(c_{{\k}\sigma})$ is the creation (annihilation)
operator for electrons of wavevector ${\k}$ and spin $\sigma$,  
$n_{i\sigma} =c_{i\sigma}^{\dagger}c_{i\sigma}$ is the number operator,
$\epsilon_{\k}^{0}=-2t\left(\cos(k_{x})+\cos(k_{y})\right)$
with $t$ being the hopping amplitude between nearest-neighbor sites, 
and $U$ is the on-site Coulomb repulsion.

We employ the dynamical cluster approximation (DCA)~\cite{hettler:dca} to study 
this model with a Quantum Monte Carlo (QMC) algorithm as the cluster solver. 
The DCA is a cluster mean-field theory which maps the original lattice  
onto a periodic cluster of size $N_c=L_c^2$ embedded in a self-consistent 
host.  Spatial correlations up to a range $L_c$ are treated explicitly, 
while those at longer length scales are described at the mean-field level.   
However the correlations in time, essential for quantum criticality,  are 
treated explicitly for all cluster sizes.  To solve the cluster problem we 
use the Hirsch-Fye QMC method~\cite{j_hirsch_86a,jarrell:dca} 
and employ the maximum entropy method~\cite{jarrell:mem} to calculate the 
real-frequency spectra.

We evaluate the results starting from the Bethe-Salpeter equation in the pairing channel:
\begin{eqnarray}
 \chi(Q)_{P,P^{\prime}} & = & \chi_{0}(Q)_P\delta_{P,P^{\prime}}      \nonumber\\
 & & +\sum_{P^{\prime\prime}}\chi(Q)_{P,P^{\prime\prime}}\Gamma(Q)_{P^{\prime\prime},P^{\prime}}\chi_{0}(Q)_{P'}
\end{eqnarray}
where $\chi$ is the dynamical susceptibility, $\chi_{0}(Q)_{P}$ [$=-G(P+Q)G(-P)$] is
 the bare susceptibility, which is constructed from $G$, the dressed one-particle Green's function, 
$\Gamma$ is the vertex function, and indices $P^{[...]}$ and external index $Q$ denote both momentum 
and frequency.  The instability of the Bethe-Salpeter equation is detected by solving
the eigenvalue equation $\Gamma\chi_{0}\phi=\lambda\phi$~\cite{Bulut93} for fixed $Q$.  By decreasing the temperature, the leading $\lambda$ 
increases to one at a temperature $T_c$ where the system undergoes a phase transition.  To 
identify which part, $\chi_{0}$ or $\Gamma$, dominates at the phase transition, we project 
them onto the d-wave pairing channel (which was found to be dominant~\cite{jarrell01, Maier2005}).  For $\chi_{0}$, we apply the $d$-wave projection 
as $\chi_{0d}(\omega)=\sum_{k}\chi_0(\omega, q=0)_{k}g_d(k)^{2}/\sum_{k}g_d(k)^{2}$, 
where $g_d(k)=(\cos( k_x)-\cos( k_y))$ is the $d$-wave form 
factor. As for the pairing strength, we employ the projection as 
$V_{d}=\sum_{k,k^{\prime}}g_d(k)\Gamma_{k,k^{\prime}} g_d(k')/\sum_{k}g_d(k)^2,$
using $\Gamma$ at the lowest Mastsubara frequency~\cite{maier}. 

To further explore the different contributions to the pairing vertex, we employ
the formally exact parquet equations to decompose it into different components~\cite{maier, parquet}.
Namely, the fully irreducible vertex $\Lambda$, the charge (S=0) particle-hole contribution, 
$\Phi_{c}$, and the spin (S=1) particle-hole contribution, $\Phi_{s}$, through: 
$\Gamma=\Lambda+\Phi_{c}+\Phi_{s}$. Similar to the previous 
expression, one can write $V_{d}=V_{d}^{\Lambda}+V_d^c+V_d^m$, where each term is the d-wave 
component of the corresponding term. Using this scheme, we
will be able to identify which component contributes the most to the d-wave pairing interaction.

\paragraph*{Results-} We use the BCS-like approximation, discussed above, to study the proximity 
of the superconducting dome to the QCP.  We take $U=6t$ ($4t=1$) on $12$ and $16$ site 
clusters large enough to see strong evidence for a QCP near doping $\delta \approx 0.15$~\cite{Vidhyadhiraja09,karlis,ehsan}.  
We explore the physics down to $T\approx 0.11J$ on the $16$ site cluster and $T\approx 0.07J$ on
the $12$-site cluster, where $J\approx 0.11$~\cite{Macridin07} is the antiferromagnetic exchange 
energy. The fermion sign problem prevents access to  lower $T$.

\begin{figure}[t]
\includegraphics*[width=8cm]{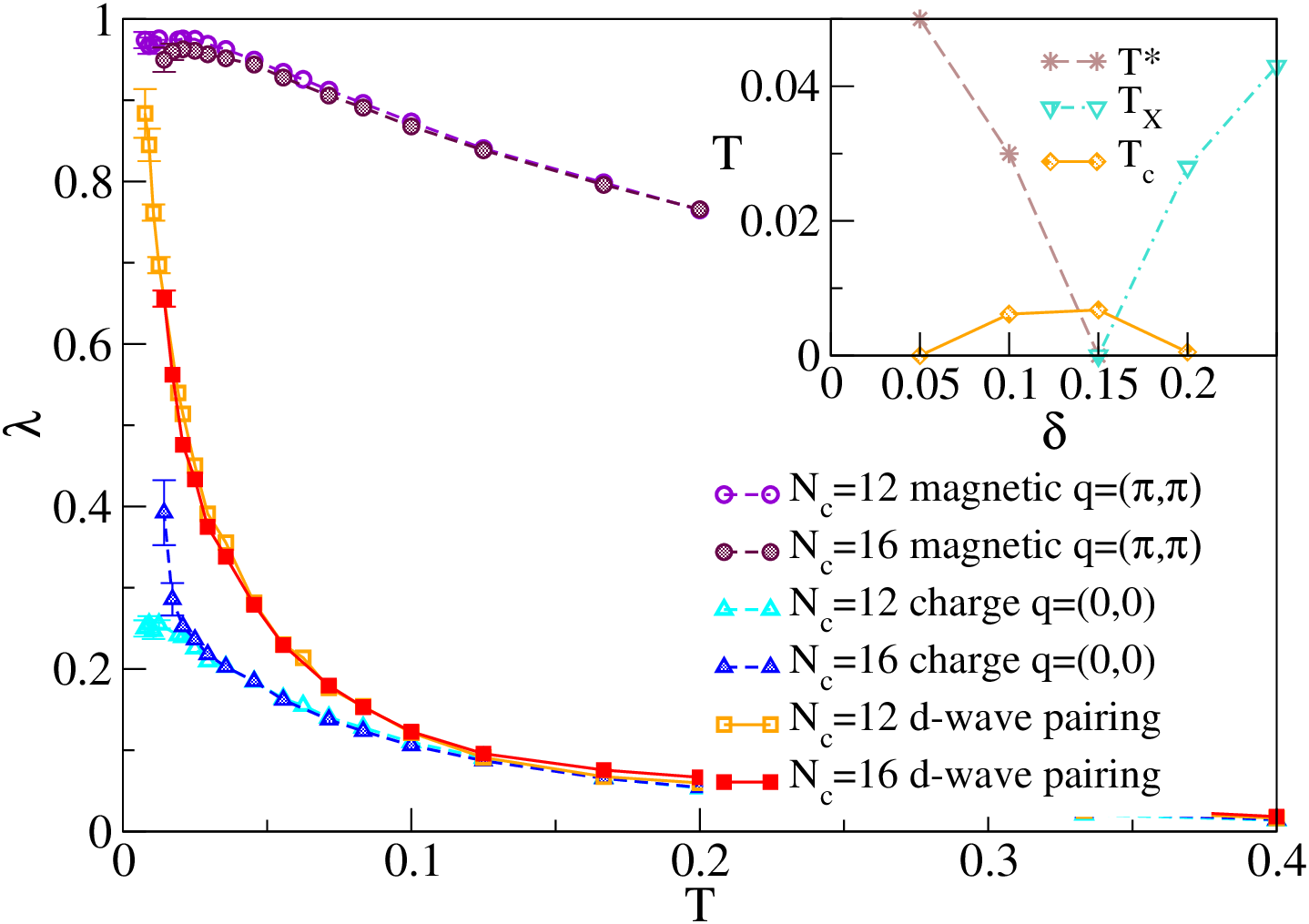}
\caption{(Color online) Plots of leading eigenvalues for different channels at the critical doping 
for $N_c=12$ and $N_c=16$ site clusters.  The inset shows the phase diagram with superconducting 
dome, pseudogap $T^*$ and FL $T_X$ temperatures from Ref.~\cite{Vidhyadhiraja09}} 
\label{fig:eigs}
\end{figure}

Fig.~\ref{fig:eigs} displays the eigenvalues of different channels (pair, charge, magnetic) at the
QC filling.  The results for the two cluster sizes are nearly identical, and the  pairing 
channel eigenvalue approaches one at low $T$, indicating a superconducting d-wave transition at roughly 
$T_c=0.007$.  However, in contrast to what was found previously~\cite{maier}, the 
$q=0$ charge eigenvalue is also strongly enhanced, particularly for the larger $N_c=16$ cluster,  
as it is expected from a QCP emerging as the terminus of a line of 
second-order critical points of charge separation transitions~\cite{ehsan}. 
 The inset shows the phase diagram, including the superconducting dome and the 
pseudogap $T^*$ and FL $T_X$ temperatures.

\begin{figure}[t]
\includegraphics*[width=8cm]{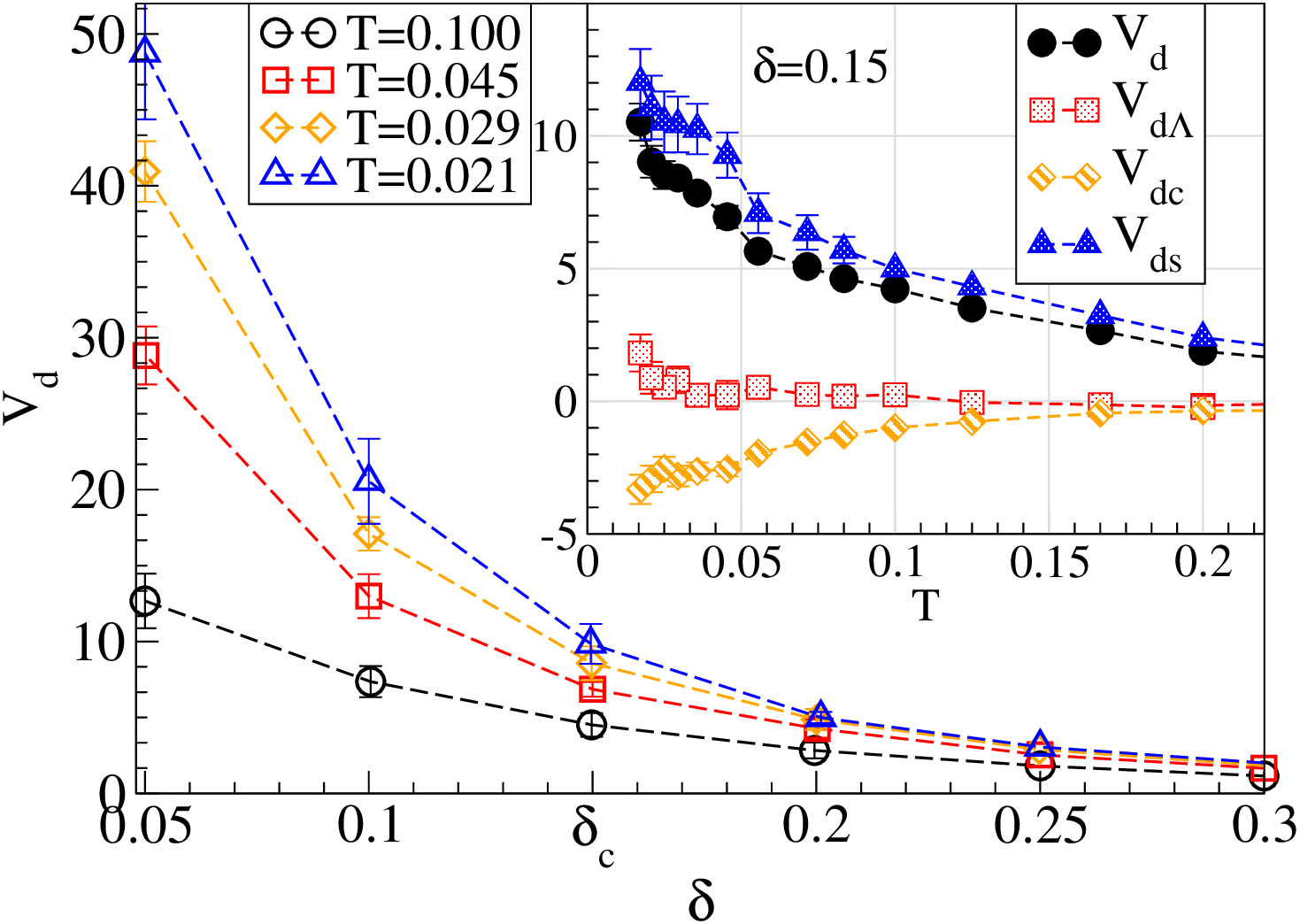}
\caption{(Color online) Plots of $V_d$, the strength of the $d$-wave pairing interaction for various 
temperatures with $U=1.5$ ($4t=1$) and $N_c=16$. $V_d$ decreases monotonically with doping, 
and shows no feature at the critical doping.  In the inset are plots of 
the contributions to $V_d$ from the charge $V_d^c$ and spin $V_d^s$ cross channels and from the 
fully irreducible vertex $V_d^\Lambda$ versus $T$ at the critical doping.  As the 
temperature is lowered, $T\ll J \approx 0.11$, the contribution to the pairing interaction from the 
spin channel is clearly dominant.} \label{fig:Vd}
\end{figure}

In Fig.~\ref{fig:Vd}, we show the strength of the $d$-wave pairing vertex $V_d$ versus 
doping for a range of temperatures.  Consistent with previous studies~\cite{maier1}, 
we find that $V_d$ falls monotonically with increasing doping. At the critical doping, 
$\delta_c=0.15$, $V_d$ shows no feature, invalidating the second scenario described above. 
The different components of $V_d$ at the critical 
doping versus temperature are shown in the inset of Fig.~\ref{fig:Vd}. 
As the QCP is approached, the pairing originates predominantly from the spin channel.  
This is similar to the result of Ref.~\cite{maier} where the pairing interaction was studied 
away from quantum criticality.   

\begin{figure}[t]
\includegraphics*[width=7cm]{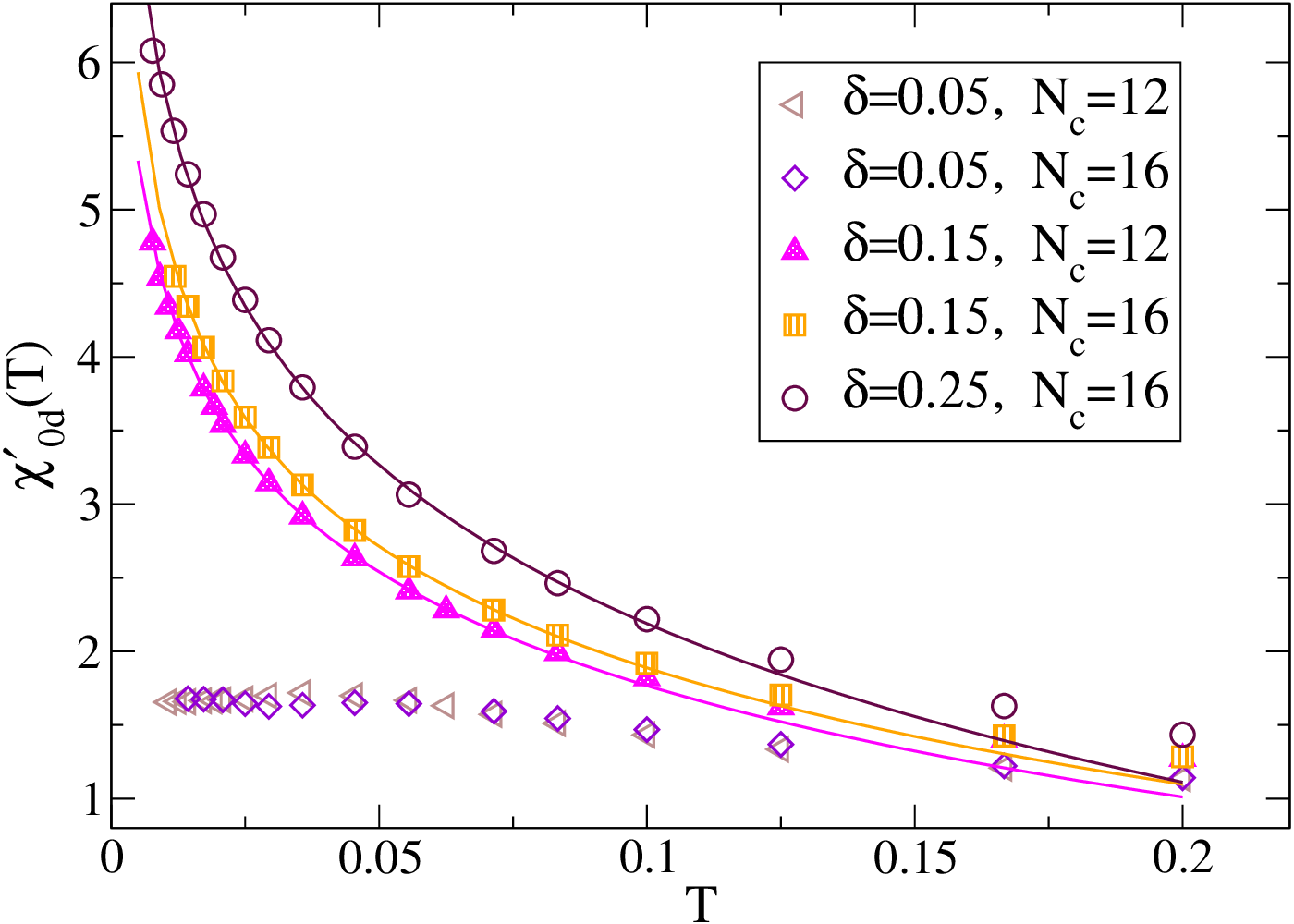}
\caption{(Color online) Plots of $\chi_{0d}'(\omega=0)$, the real part of the bare $d$-wave pairing 
susceptibility, at zero frequency vs.\ temperature at three characteristic dopings. The solid lines
are fits to $\chi_{0d}'(\omega=0) = B/\sqrt{T} + A\ln(\omega_c/T)$ for $T<J$.  In the 
underdoped case ($\delta=0.05$), $\chi_{0d}'(\omega=0)$ does not grow with decreasing temperature. At 
the critical doping ($\delta=\delta_c=0.15$), $\chi_{0d}'(\omega=0)$ shows power-law behavior with
$B=0.04$ for the 12 site, and $B=0.09$ for the 16-site clusters (in both $A=1.04$ and $\omega_c=0.5$). 
In the overdoped region ($\delta=0.25$), a log divergence is found, with $B=0$ obtained from the fit.
} \label{fig:chi_0d_new}
\end{figure}

In contrast, the bare $d$-wave pairing susceptibility $\chi_{0d}$ exhibits significantly 
different features near and away from the QCP.  As shown in Fig.~\ref{fig:chi_0d_new}, in the 
underdoped region (typically $\delta=0.05$), % which is also known as the pseudogap or NFL region, 
the bare $d$-wave pairing susceptibility $\chi_{0d}'(\omega=0)$ saturates at low temperatures. 
However, at the critical doping, it diverges quickly with decreasing temperature, roughly following 
the power-law behavior $1/\sqrt{T}$, while in the overdoped or FL region it displays a log 
divergence.

\begin{figure}[t]
\includegraphics*[width=8cm]{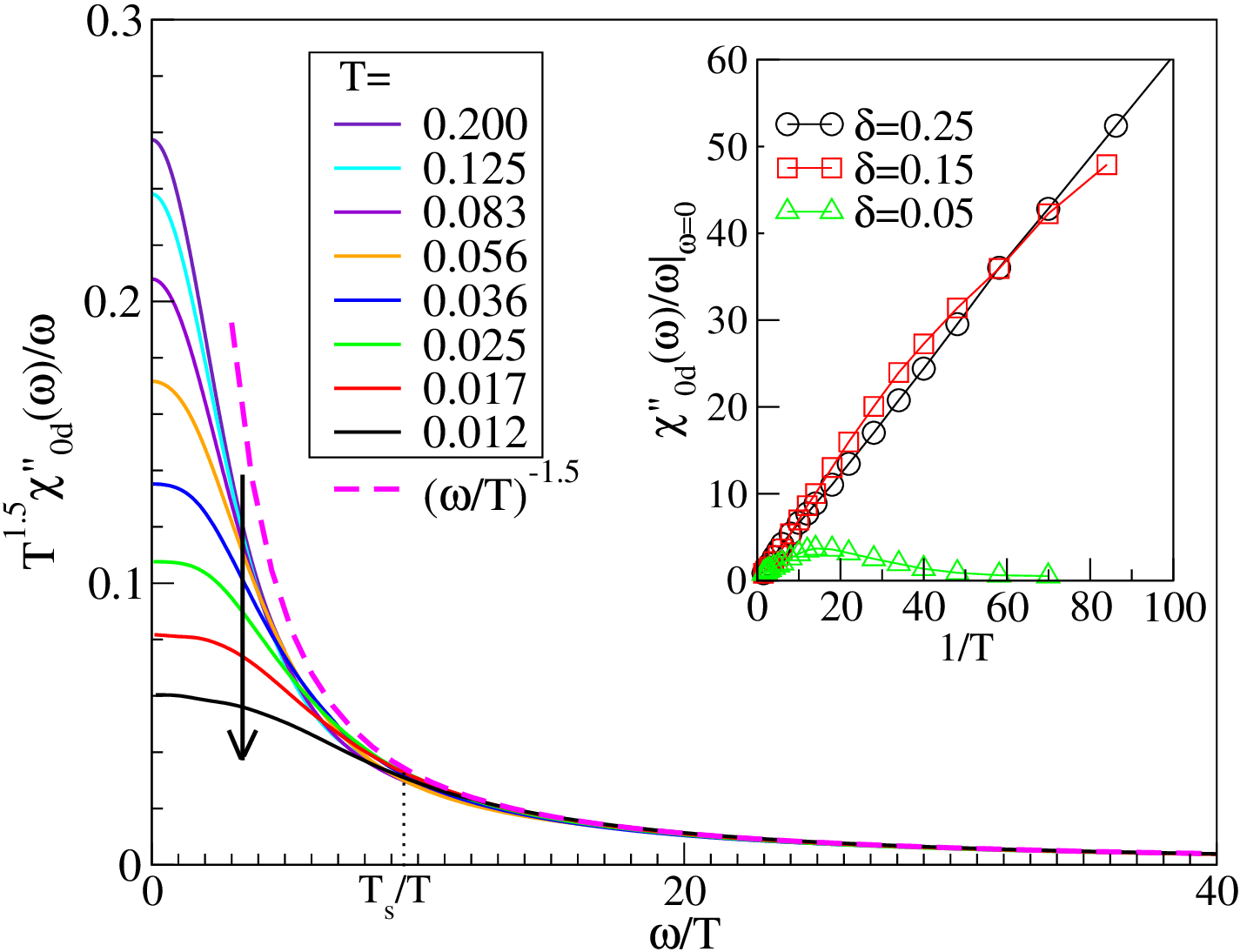}
\caption{(Color online) Plots of $T^{1.5}\chi_{0d}''(\omega)/\omega$ versus $\omega/T$ at the QC 
doping ($\delta=0.15$) for $N_c=16$.    The arrow denotes the direction of decreasing temperature. 
The curves coincide for $\omega/T>9\approx (4t/J)$ defining a scaling function $H(\omega/T)$, 
corresponding to a contribution to 
$\chi_{0d}^\prime (T) = \frac{1}{\pi}\int d\omega \chi_{0d}''(w)/w \propto 1/\sqrt{T}$ as found 
in Fig.~\ref{fig:chi_0d_new}.  For $\omega/T>9\approx (4t/J)$, $H(\omega/T) \approx (\omega/T)^{-1.5}$
(dashed line).  On the x-axis, we add the label $T_s/T\approx (4t/J)$, where $T_s$ 
represents the energy scale where curves start deviating from $H$.  The inset shows the unscaled 
zero-frequency result 
$\left . \chi_{0d}''(\omega)/\omega\right |_{\omega=0}$ plotted versus inverse temperature.
}
\label{fig:weta}
\end{figure}

To better understand the temperature-dependence of 
$\chi_{0d}'(\omega=0)$ at the QC doping, we looked 
into %its imaginary part divided by frequency scaled as 
$T^{1.5} \chi_{0d}''(\omega)/\omega$  and plotted it
versus $\omega/T$ in Fig.~\ref{fig:weta}.   When scaled this way, the curves from different temperatures 
fall on each other such that $T^{1.5}\chi''_{0d}(\omega)/\omega= H(\omega/T) \approx (\omega/T)^{-1.5}$ 
for $\omega/T \agt 9 \approx 4t/J$.   For $0<\omega/T<4t/J$, the curves deviate from the scaling function 
$H(x)$ and show nearly BCS behavior, with $\left . \chi_{0d}''(\omega)/\omega\right |_{\omega=0}$ which 
is weakly sublinear in $1/T$ as shown in the inset.  The curves away from the critical doping (not displayed) 
do not show such a collapse.  In the underdoped region ($\delta=0.05$) at low frequencies, 
$\chi_{0d}''(\omega)/\omega$ goes to zero with decreasing temperature (inset). In the FL region 
($\delta=0.25$) $\chi_{0d}''(\omega)/\omega$ develops a narrow peak at low $\omega$ of width 
$\omega \approx T_X$ and height $\propto 1/T$ as shown in the inset.

\paragraph*{Discussion-}  

$\chi''_{0d}(\omega)/\omega$ reveals details about how the instability takes place.   The overlapping 
curves found at the QC filling contribute a term $ T^{-1.5} H(\omega/T) $ to $\chi_{0d}''(w)/w$ or 
$\chi_{0d}^\prime (T) \propto 1/\sqrt{T}$ as found in Fig.~\ref{fig:chi_0d_new}.  There is also a 
component which does not scale, especially at low frequencies.  In fact, $\chi_{0d}''(\omega)/\omega$ 
at zero frequency increases more slowly than $1/T$ as expected for a FL.  From this sublinear character, 
we infer that the contribution of the non-scaling part of $\chi''_{0d}(\omega)/\omega$ to the divergence 
of $\chi_{0d}^\prime (T)$ is weaker than BCS and may cause us to overestimate $A$ and underestimate $B$ 
in the fits performed at the critical doping in Fig.~\ref{fig:chi_0d_new}. In addition, if $H(0)$ is 
finite, it would contribute a term to $\chi_{0d}^\prime (T)$ that increases like $1/T^{1.5}$, so $H(0)=0$.  
From Eq.~\ref{eq:chipp} we see that the contribution to $\chi_{0d}''(\omega)/\omega$ at small $\omega$ 
comes only from states near the Fermi surface.  $H(0)=0$ would indicate that the enhanced pairing 
associated with $\chi_{0d}^\prime (T) \propto 1/\sqrt{T}$ is due to higher energy states.  The vanishing 
of $\chi''_{0d}(\omega)/\omega$ in the pseudogap region ($\delta=0.05$)
for small frequency  when $T\rightarrow 0$ indicates that around the Fermi surface, the dressed 
particles do not respond to a pair field.  Or, perhaps more correctly, none are available for pairing
due to the pseudogap depletion of electron states around the Fermi surface.  Thus, even the strong 
$d$-wave interaction, seen in Fig.~\ref{fig:Vd}, is unable to drive the system into a superconducting 
phase.  In the overdoped region, $\chi''_{0d}(\omega)/\omega$ displays conventional
FL behavior for $T<T_X$, and the vanishing $V_d$ suppresses $T_c$.

Together, the results for  $\chi_{0d}$ and  $V_d$ shed light on the shape of the superconducting dome in 
the phase diagram found previously~\cite{ehsan}. With increasing doping, the pairing vertex $V_d$ falls 
monotonically. On the other hand, $\chi'_{0d}(T)$ is strongly suppressed in the low doping or pseudogap 
region and enhanced at the critical and higher doping. These facts alone could lead to a superconducting 
dome.  Futhermore, the additional algebraic divergence of $\chi'_{0d}(T)$ seen in Fig.~\ref{fig:chi_0d_new} 
causes the superconductivity to be enhanced even more strongly near the QCP where one might expect 
$T_c \propto \left(V_d B\right)^{2}$, with $B=\frac{1}{\pi} \int dx H(x)$, 
compared to the conventional BCS form in the FL region.

Similar to the scenario for cuprate superconductivity suggested by Castellani {\em{et al.}}~\cite{rome}, 
we find that the superconducting dome is due to charge fluctuations adjacent to the QCP related to charge 
ordering. However, we differ in that we find the pairing in this region is due to an algebraic temperature dependence of the bare susceptibility $\chi_{0d}$ rather than an enhanced $d$-wave pairing vertex $V_d$, and that this pairing interaction is dominated by the spin channel.

Our observation in the Hubbard model offers an experimental accessible variant of She-Zaanen's QCBCS.
We use the bare pairing susceptibility
$\chi_0$ while She-Zaanen use the full $\chi$, which includes all the effects of quantum criticality but not the
correction from the pairing vertex (the pairing glue is added separately). This decomposition is not possible
in numerical calculations or experiments since both quantum criticality and pairing originate from
the Coulomb interaction.  However, the effect of quantum criticality already shows up in the one-particle
quantities, % such as the self energy and single-particle spectra, in that 
and the spectra have different behaviors
for the three regions around the superconducting dome. She-Zaanen assume that 
$\chi''(\omega) \propto 1/\omega^{\alpha}$ for $T_s<\omega<\omega_c$, where $\omega_c$ is an upper cutoff,
and that it is irrelevant $(\alpha<0)$, marginal ($\alpha=0$), or relevant ($\alpha>0$), respectively 
in the pseudo gap region, FL region and QCP vincity.  We find the same behavior in $\chi_0$ 
and we have the further observation that near the QCP $T_s\approx (4t/J)T$ and $\alpha=0.5$. 

Experiments combining angle-resolved photo emission (ARPES) and inverse photo emission results, with an energy 
resolution of roughly $J$, could be used to construct $\chi_{0d}$ and explore power law scaling at the critical doping.  
Since the energy resolution of ARPES is much better than inverse photo emission, it is also interesting to study 
$\chi''_{0d}(\omega)/\omega |_{\omega=0}$, which only requires ARPES data, but not inverse photo emission.

\paragraph*{Conclusion-}
Using the DCA, we investigate the $d$-wave pairing instability in the two-dimensional Hubbard model near 
critical doping.  We find that the pairing interaction remains dominated by the spin channel and is 
not enhanced near the critical doping.   However, we find a power-law divergence of the bare pairing 
susceptibility at the critical doping, replacing the conventional BCS logarithmic behavior.  We interpret 
this behavior by studying the dynamic bare pairing susceptibility which has a part that scales like 
$\chi''_{0d}(\omega)/\omega \sim T^{-1.5} H(\omega/T)$, where $H(\omega/T)$ is a universal function.
Apparently, the NFL character of the QCP yields an electronic system that is far more susceptible 
to d-wave pairing than the FL and pseudogap regions.  We also suggest possible experimental approaches 
to exploit this interesting behavior.

\paragraph*{Acknowledgments-}   We would like to thank 
F.~Assaad,
I.~Vekhter
and
E. W.~Plummer
for useful conversations.  This research was supported by NSF DMR-0706379 and OISE-0952300.  This research 
used resources of the National Center for Computational Sciences (Oak Ridge National Lab), which is 
supported by the DOE Office of Science under Contract No.\ DE-AC05-00OR22725.
J.-H. She and J. Zaanen are supported by the Nederlandse Organisatie voor Wetenschappelijk Onderzoek (NWO) via a
Spinoza grant.

\end{document}